\documentclass[%
aps,
prl,
twocolumn,
reprint,
superscriptaddress,
showpacs,preprintnumbers,
 amsmath,amssymb,
 aps,
floatfix,
]{revtex4-1}

\usepackage{lipsum}
\usepackage{hyperref}
\usepackage{graphicx} 
\usepackage{dcolumn}  
\usepackage{graphicx} 
\usepackage{dcolumn}  
\usepackage{xspace}
\usepackage{amsmath} 
\usepackage{cleveref}
\usepackage{subfigure}
\usepackage{booktabs}
\usepackage{longtable}
\usepackage{dcolumn}

\usepackage[usenames,dvipsnames,svgnames,table]{xcolor}

\hypersetup{colorlinks=true, citecolor=Blue, urlcolor=Blue, linkcolor=Blue}

\graphicspath{{ps}}

\renewcommand{\arraystretch}{1.0}

\begin{document}

\vspace*{0.2cm}

\preprint{\vbox{ \hbox{   }
						\hbox{Belle Preprint 2020-14}
						\hbox{KEK Preprint   2020-16}
}}

\newcommand{\Bto}[1]{
	\ifnum #1 = 521343 \ensuremath{{B^+ \to K^+ e^+ e^-} \xspace}\fi
	\ifnum #1 = 521347 \ensuremath{{B^+ \to K^+ \mu^+ \mu^-} \xspace}\fi
	\ifnum #1 = 521345 \ensuremath{B^+ \to K^{\ast +} e^+ e^- \xspace}\fi
	\ifnum #1 = 521349 \ensuremath{B^+ \to K^{\ast +} \mu^+ \mu^- \xspace}\fi
	\ifnum #1 = 511332 \ensuremath{B^0 \to K^0 e^+ e^- \xspace}\fi
	\ifnum #1 = 511336 \ensuremath{B^0 \to K^0 \mu^+ \mu^- \xspace}\fi
	\ifnum #1 = 511335 \ensuremath{B^0 \to K^{\ast 0} e^+ e^- \xspace}\fi
	\ifnum #1 = 511339 \ensuremath{{B^0 \to K^{\ast 0}  \mu^+ \mu^-} \xspace}\fi
}

\newcommand{\gev}{\ensuremath{\mathrm{GeV}}\xspace}
\newcommand{\bmn}{$B$ meson\xspace}
\newcommand{\bms}{$B$ mesons\xspace}
\newcommand{\gevc}{\ensuremath{\mathrm{GeV}/c^2}\xspace}
\newcommand{\gevsq}{\ensuremath{~\mathrm{GeV}^2/c^4}\xspace}
\newcommand{\mbc}{\ensuremath{M_\mathrm{bc}}\xspace}
\newcommand{\bkstll}{$B \to K^{\ast} \ell^+ \ell^-$\xspace}
\newcommand*{\factorh}{0.48}
\newcommand{\bsll}{$b \to s \ell^+ \ell^-$\xspace}
\newcommand{\fitcomponentstwo}{Combinatorial (dashed blue), signal (red filled), charmonium (dashed green), peaking (purple dotted), and total (solid) fit distributions are superimposed on data (points  with error bars).}

\title{Test of lepton-flavor universality  in  ${B\to K^\ast\ell^+\ell^-}$  decays at Belle}

\noaffiliation
\affiliation{University of the Basque Country UPV/EHU, 48080 Bilbao}
\affiliation{University of Bonn, 53115 Bonn}
\affiliation{Brookhaven National Laboratory, Upton, New York 11973}
\affiliation{Budker Institute of Nuclear Physics SB RAS, Novosibirsk 630090}
\affiliation{University of Cincinnati, Cincinnati, Ohio 45221}
\affiliation{Deutsches Elektronen--Synchrotron, 22607 Hamburg}
\affiliation{University of Florida, Gainesville, Florida 32611}
\affiliation{Department of Physics, Fu Jen Catholic University, Taipei 24205}
\affiliation{Key Laboratory of Nuclear Physics and Ion-beam Application (MOE) and Institute of Modern Physics, Fudan University, Shanghai 200443}
\affiliation{Justus-Liebig-Universit\"at Gie\ss{}en, 35392 Gie\ss{}en}
\affiliation{SOKENDAI (The Graduate University for Advanced Studies), Hayama 240-0193}
\affiliation{Department of Physics and Institute of Natural Sciences, Hanyang University, Seoul 04763}
\affiliation{University of Hawaii, Honolulu, Hawaii 96822}
\affiliation{High Energy Accelerator Research Organization (KEK), Tsukuba 305-0801}
\affiliation{J-PARC Branch, KEK Theory Center, High Energy Accelerator Research Organization (KEK), Tsukuba 305-0801}
\affiliation{Higher School of Economics (HSE), Moscow 101000}
\affiliation{Forschungszentrum J\"{u}lich, 52425 J\"{u}lich}
\affiliation{Hiroshima Institute of Technology, Hiroshima 731-5193}
\affiliation{IKERBASQUE, Basque Foundation for Science, 48013 Bilbao}
\affiliation{Indian Institute of Science Education and Research Mohali, SAS Nagar, 140306}
\affiliation{Indian Institute of Technology Hyderabad, Telangana 502285}
\affiliation{Indian Institute of Technology Madras, Chennai 600036}
\affiliation{Indiana University, Bloomington, Indiana 47408}
\affiliation{Institute of High Energy Physics, Chinese Academy of Sciences, Beijing 100049}
\affiliation{Institute of High Energy Physics, Vienna 1050}
\affiliation{Institute for High Energy Physics, Protvino 142281}
\affiliation{INFN - Sezione di Napoli, 80126 Napoli}
\affiliation{INFN - Sezione di Torino, 10125 Torino}
\affiliation{Advanced Science Research Center, Japan Atomic Energy Agency, Naka 319-1195}
\affiliation{J. Stefan Institute, 1000 Ljubljana}
\affiliation{Institut f\"ur Experimentelle Teilchenphysik, Karlsruher Institut f\"ur Technologie, 76131 Karlsruhe}
\affiliation{Kavli Institute for the Physics and Mathematics of the Universe (WPI), University of Tokyo, Kashiwa 277-8583}
\affiliation{Kennesaw State University, Kennesaw, Georgia 30144}
\affiliation{Korea Institute of Science and Technology Information, Daejeon 34141}
\affiliation{Korea University, Seoul 02841}
\affiliation{Kyoto Sangyo University, Kyoto 603-8555}
\affiliation{Kyungpook National University, Daegu 41566}
\affiliation{Universit\'{e} Paris-Saclay, CNRS/IN2P3, IJCLab, 91405 Orsay}
\affiliation{\'Ecole Polytechnique F\'ed\'erale de Lausanne (EPFL), Lausanne 1015}
\affiliation{P.N. Lebedev Physical Institute of the Russian Academy of Sciences, Moscow 119991}
\affiliation{Faculty of Mathematics and Physics, University of Ljubljana, 1000 Ljubljana}
\affiliation{Ludwig Maximilians University, 80539 Munich}
\affiliation{Luther College, Decorah, Iowa 52101}
\affiliation{University of Maribor, 2000 Maribor}
\affiliation{School of Physics, University of Melbourne, Victoria 3010}
\affiliation{University of Mississippi, University, Mississippi 38677}
\affiliation{University of Miyazaki, Miyazaki 889-2192}
\affiliation{Moscow Physical Engineering Institute, Moscow 115409}
\affiliation{Graduate School of Science, Nagoya University, Nagoya 464-8602}
\affiliation{Kobayashi-Maskawa Institute, Nagoya University, Nagoya 464-8602}
\affiliation{Universit\`{a} di Napoli Federico II, 80126 Napoli}
\affiliation{Nara Women's University, Nara 630-8506}
\affiliation{National Central University, Chung-li 32054}
\affiliation{National United University, Miao Li 36003}
\affiliation{Department of Physics, National Taiwan University, Taipei 10617}
\affiliation{H. Niewodniczanski Institute of Nuclear Physics, Krakow 31-342}
\affiliation{Nippon Dental University, Niigata 951-8580}
\affiliation{Niigata University, Niigata 950-2181}
\affiliation{Novosibirsk State University, Novosibirsk 630090}
\affiliation{Osaka City University, Osaka 558-8585}
\affiliation{Pacific Northwest National Laboratory, Richland, Washington 99352}
\affiliation{Panjab University, Chandigarh 160014}
\affiliation{Peking University, Beijing 100871}
\affiliation{University of Pittsburgh, Pittsburgh, Pennsylvania 15260}
\affiliation{Punjab Agricultural University, Ludhiana 141004}
\affiliation{Research Center for Nuclear Physics, Osaka University, Osaka 567-0047}
\affiliation{Department of Modern Physics and State Key Laboratory of Particle Detection and Electronics, University of Science and Technology of China, Hefei 230026}
\affiliation{Seoul National University, Seoul 08826}
\affiliation{Showa Pharmaceutical University, Tokyo 194-8543}
\affiliation{Soochow University, Suzhou 215006}
\affiliation{Soongsil University, Seoul 06978}
\affiliation{Stefan Meyer Institute for Subatomic Physics, Vienna 1090}
\affiliation{Sungkyunkwan University, Suwon 16419}
\affiliation{School of Physics, University of Sydney, New South Wales 2006}
\affiliation{Department of Physics, Faculty of Science, University of Tabuk, Tabuk 71451}
\affiliation{Tata Institute of Fundamental Research, Mumbai 400005}
\affiliation{Department of Physics, Technische Universit\"at M\"unchen, 85748 Garching}
\affiliation{School of Physics and Astronomy, Tel Aviv University, Tel Aviv 69978}
\affiliation{Toho University, Funabashi 274-8510}
\affiliation{Earthquake Research Institute, University of Tokyo, Tokyo 113-0032}
\affiliation{Department of Physics, University of Tokyo, Tokyo 113-0033}
\affiliation{Tokyo Institute of Technology, Tokyo 152-8550}
\affiliation{Tokyo Metropolitan University, Tokyo 192-0397}
\affiliation{Virginia Polytechnic Institute and State University, Blacksburg, Virginia 24061}
\affiliation{Wayne State University, Detroit, Michigan 48202}
\affiliation{Yamagata University, Yamagata 990-8560}
\affiliation{Yonsei University, Seoul 03722}

\author{S.~Wehle}\affiliation{Deutsches Elektronen--Synchrotron, 22607 Hamburg} 

\author{I.~Adachi}\affiliation{High Energy Accelerator Research Organization (KEK), Tsukuba 305-0801}\affiliation{SOKENDAI (The Graduate University for Advanced Studies), Hayama 240-0193} 
\author{K.~Adamczyk}\affiliation{H. Niewodniczanski Institute of Nuclear Physics, Krakow 31-342} 
\author{H.~Aihara}\affiliation{Department of Physics, University of Tokyo, Tokyo 113-0033} 
\author{D.~M.~Asner}\affiliation{Brookhaven National Laboratory, Upton, New York 11973} 
\author{H.~Atmacan}\affiliation{University of Cincinnati, Cincinnati, Ohio 45221} 
\author{V.~Aulchenko}\affiliation{Budker Institute of Nuclear Physics SB RAS, Novosibirsk 630090}\affiliation{Novosibirsk State University, Novosibirsk 630090} 
\author{T.~Aushev}\affiliation{Higher School of Economics (HSE), Moscow 101000} 
\author{R.~Ayad}\affiliation{Department of Physics, Faculty of Science, University of Tabuk, Tabuk 71451} 
\author{V.~Babu}\affiliation{Deutsches Elektronen--Synchrotron, 22607 Hamburg} 
% ...
\collaboration{The Belle Collaboration}

\noaffiliation

\begin{abstract}
	We present a measurement of $R_{K^{\ast}}$, the branching fraction ratio ${{\cal B}(B\to K^\ast \mu^+ \mu^-)}$/ ${{\cal B}(B\to K^\ast e^+ e^-)}$, for both charged   and neutral $B$ mesons. 
	The ratio for the charged case, $R_{K{^{\ast +}}}$, is the first measurement ever performed. 
	In addition, we report absolute branching fractions for the
	individual modes in bins of the squared dilepton invariant mass, $q^2$.
	The analysis is based on a data sample of $711~\mathrm{fb}^{-1}$,  containing $772\times 10^{6}$ $B\bar B$ events, recorded at the $\Upsilon(4S)$ resonance with the Belle detector at the KEKB asymmetric-energy $e^+e^-$ collider.
	The obtained results are consistent  with Standard Model expectations.
\end{abstract}


\maketitle

\clearpage



In the Standard Model (SM), the coupling of gauge bosons to leptons is independent of lepton-flavor, a concept known as lepton-flavor universality (LFU). 
Therefore, experimental tests of LFU are  excellent probes for New Physics (NP). 
In this Letter, we present a test of LFU in $B\to K^\ast\ell^+\ell^-$ decays, where $\ell$ is either $e$ or $\mu$.
These decays have been studied by several experiments, and some results suggest an intriguing possibility that the underlying $b \to s \ell^+ \ell^-$  transition may be affected by physics beyond the SM in modes involving muons \cite{Wehle:2016yoi,Aaij:2014ora,Aaij:2015oid,Aaij:2017vbb,Lees:2012tva,Wei:2009zv}.
The ratio of branching fractions,
\begin{equation}		
	R_{K^\ast} = \frac{{\cal B}(B\to K^\ast \mu^+ \mu^-)}{{\cal B}(B\to K^\ast e^+ e^-)}, 
	\label{eq:rkst}
\end{equation}
is well suited  to test  LFU \cite{Hiller:2014ula}.
The theoretical predictions for $R_{K^\ast}$ are robust \cite{Hiller:2014ula,Bobeth:2007dw,Bordone:2016gaq}, as uncertainties related to form factors cancel out in the ratio.
This observable is expected to be  close to unity in the SM.

For this measurement, we reconstruct the decay channels \Bto{511339},  \Bto{521349}, \Bto{511335}, and   \Bto{521345}.
The  $K^\ast$ meson is reconstructed in the $K^+\pi^-$,  $K^+\pi^0$, and  $K^0_S\pi^+$ decay modes. 
The inclusion of charge-conjugate states is implied throughout this paper.
Compared to the previous  analysis \cite{Wei:2009zv}, the full $\Upsilon(4S)$ data sample   containing $772\times 10^{6}$ $B\bar B$ events, recorded with the Belle detector \cite{BelleDetektor} at the KEKB asymmetric-energy $e^+e^-$ collider  \cite{kekb}, is used.

Belle  is a large-solid-angle magnetic
spectrometer that consists of a silicon vertex detector,
a 50-layer central drift chamber (CDC), an array of
aerogel threshold Cherenkov counters (ACC),  
a barrel-like arrangement of time-of-flight
scintillation counters (TOF), and an electromagnetic calorimeter
comprised of CsI(Tl) crystals (ECL). 
All these components are located inside a superconducting solenoid coil that provides a 1.5~T
magnetic field.  
An iron flux return placed outside of the coil is instrumented with resistive plate chambers to detect $K_L^0$ mesons and muons (KLM).  

The  analysis is validated and optimized with simulated  Monte Carlo (MC) data samples, from which also  the selection efficiencies are derived. 
The   EvtGen \cite{evtgen} and PYTHIA \cite{pythia} packages are used to generate  decay chains, where the final-state-radiation effect is incorporated with PHOTOS \cite{photos}.
The  detector response is simulated with  GEANT3 \cite{geant}.


All  tracks, except for those from $K_S^0$ decays, need to satisfy requirements on their impact parameter  with respect to the  interaction point along the $z$ axis ($|dz| <5.0~\textrm{cm}$)  and in the transverse $x$-$y$ plane   ($|dr| <1.0~\textrm{cm}$).
The $z$ axis is in the direction opposite to that of the $e^{+}$ beam. 
We calculate a particle identification (PID) likelihood for each track using  energy loss  in the CDC, information from the TOF, number of the photoelectrons from the ACC, the transverse shower shape and energy  in the ECL, and hit information from the KLM.
Electrons are identified using the likelihood ratio  ${\cal P}_{e}=L_e/(L_e+L_\pi)$, where $L_{i}$ is the PID  likelihood for the particle type $i$.
Charged tracks satisfying ${\cal P}_{e} > 0.9$ are accepted as electron candidates.
Energy losses due to bremsstrahlung are recovered by adding the momenta of photons to that of the electron's momentum if they lie within $0.05$ rad of the initial track direction.
Tracks are selected as muon candidates   if they satisfy  ${\cal P}_{\mu} > 0.9$, where ${\cal P}_{\mu}$ is the analogous likelihood ratio for muons.
For electron (muon) candidates we require the momentum to be greater than  0.4 (0.7) $\gev/c$ so that they can reach the ECL (KLM), which improves the PID.
These requirements select electron (muon) candidates with an efficiency greater than 86\% (92\%) while rejecting more than 99\% of pions.
Charged kaons are distinguished from pions (and vice versa) by requiring the likelihood ratio ${\cal P}_{K}=L_K/(L_K+L_\pi)$ to be greater than 0.1 (smaller than 0.9). 
This requirement retains more than 99\% of kaons (pions) while reducing the misidentification rate of  pions (kaons) by 94\% (86\%).

The $K_S^0$ candidates  are reconstructed with an efficiency of $74\%$ from two  oppositely charged tracks (treated as pions) by applying selection criteria on their    invariant mass  and  vertex-fit quality \cite{Chen:2005dra}.
We reconstruct $\pi^0$ candidates from photon pairs, where each photon is required to have an energy greater than $30~\mathrm{MeV}$. Furthermore, the invariant mass of the photon pair is required to be in the  $[115, 153]~\mathrm{MeV}/c^2$ range, which corresponds to  approximately $\pm 4$ times the $\pi^0$ reconstructed-mass resolution.
We form $K^*$ candidates  from $K^+\pi^-$,  $K^+\pi^0$, and  $K^0_S\pi^+$ combinations with  an invariant mass lying  in the range
$[0.6, 1.4]~\textrm{GeV}/c^2$.
We also apply a requirement on the $K^*$ vertex-fit quality to reduce background.
The $K^{\ast}$ candidates are  combined  with two oppositely charged leptons to form \bmn candidates. 


The dominant background is due to incorrect combinations of tracks.
This combinatorial background  is  suppressed by applying requirements on the  beam-energy-constrained  mass, $ {M_\textrm{bc} =  \sqrt{E^2_{\mathrm{beam}}/c^4 - |\vec p_B|^2/c^2}}$, and the energy difference, $\Delta E =  E_B - E_{\mathrm{beam}}$, where $E_{\mathrm{beam}}$ is the beam energy, and $E_B$ and $\vec p_B$ are the energy and  momentum, respectively, of the reconstructed $B$-meson candidate.
All of these quantities are calculated in the center-of-mass frame.
Correctly reconstructed signal events peak near the $B$-meson mass \cite{pdg} in \mbc  and at zero in $\Delta E$.
The $\Delta E$ distribution is wider for electron modes as some bremsstrahlung photons are not reconstructed. 
We retain $B$ decay candidates that satisfy $ 5.22~\gevc < M_\mathrm{bc}  <~5.30~\mathrm{GeV}/c^2 $ and $ -0.10 \ (-0.05)~\gev <\Delta E < ~0.05~\mathrm{GeV}$     in the electron (muon) mode.

Large irreducible background contributions  arise in the $\Delta E$ and $M_{\rm bc}$ distributions  from the decays $B\to J/\psi K^{\ast} $ and $B\to \psi(2S) K^{\ast} $, where the charmonium states further decay into two leptons. 
We veto this background  by rejecting candidates with $-0.25~ (-0.15)~\mathrm{GeV}/c^2 <  M_{\ell\ell} - m_{J/\psi}< 0.08~\mathrm{GeV}/c^2$ and $-0.20 ~ (-0.10)~\mathrm{GeV}/c^2 <  M_{\ell\ell} - m_{\psi(2S)}< 0.08~\mathrm{GeV}/c^2$ for the electron (muon) channel.
In the electron case, the veto is applied twice: before and after the bremsstrahlung-recovery treatment. 
This is done to prevent charmonium backgrounds from shifting out of the veto region when an incorrect photon is combined with the electron.

A multivariate analysis technique is developed to suppress combinatorial background.
A dedicated neural network classifier is trained with MC samples to identify each particle type used in the decay chain, from which a signal probability is calculated for each candidate.
The neural networks dedicated to identifying the particles  $e^\pm, \mu^\pm, K^\pm$, $K_S^{0}$, $\pi^0$, and $\pi^\pm$ are identical to those used in Ref.~\cite{fullrecon}.
The  networks  for $K^\ast$ selection use  input variables related to the $K^\ast$ daughter particles. 
Most of the discrimination of the $K^*$ selection comes from vertex-fit information, decay-product  neural-network outputs,  and momenta of the decay products.
The final signal selection is performed with a dedicated neural network for each $B$ decay channel.
The inputs to these $B$-decay classifiers include  event-shape variables (modified Fox-Wolfram moments \cite{ksfwm}), vertex-fit information, and kinematic variables such as the reconstructed mass of the $K^\ast$ and the angle between its momentum vector and the initial direction extracted from the vertex fit.
The most discriminating of these input variables are $\Delta E$, the reconstructed $K^\ast$ mass, the product of the network outputs for all daughter particles, and the distance between the two leptons projected onto the $z$ axis as derived from a fit to the $B$-decay vertex. 
The final selection requirement on the $B$-decay classifier output value is optimized  by maximizing a figure of merit, $n_\mathrm{s}/\sqrt{n_\mathrm{s} + n_\mathrm{b}}$,  where $n_\mathrm{s}$  ($n_\mathrm{b}$)  is the expected number of signal  (background) events calculated from MC samples in the region $M_{\rm bc} > 5.27~\gev/c^2$.

Less than 2\% of events contain multiple $B$  candidates. 
In such cases,  we choose the one with the highest signal probability, estimated from the neural network output values.
This procedure selects the correct $B$ candidate with an efficiency between 82\% and 92\%, depending on the channel.
Individual decay channels in these samples are normalized according to the  branching fraction values  reported in Ref.  \cite{pdg}.


We extract signal yields  in various regions of the squared dilepton invariant mass, $q^2$, using an unbinned extended maximum-likelihood fit to the  \mbc  distribution of  \bkstll candidates.
We consider four different components in the likelihood fit.
First, a signal component is parametrized by a Crystal Ball function \cite{CBshape}, with the shape parameters determined from  $B\to J/\psi K^\ast$ candidates that fail the $J/\psi$ veto in data.
Second, a combinatorial background component  is described by an  ARGUS function \cite{ARGUSshape}. 
Third, there is a component from events in which  charmonium  decays  pass the veto when they are misreconstructed; for example,  when the bremsstrahlung recovery fails to detect photons. 
This background component is studied using an MC sample with 100 times higher statistics than that expected from the charmonium decays in the data sample.
The shape of the charmonium  background is determined via kernel density estimation (KDE) \cite{kde}.
Lastly, a peaking background component  from double misidentification of particle type, where two particles have been assigned the wrong hypothesis such as $B\to K^\ast \pi^+ \pi^-$, is studied  using   MC samples, with the shape parametrized via KDE.
As the expected yields of charmonium and double-mis-identification backgrounds  are small, their yields are fixed in the fit to values obtained from MC  simulation.


The determination of signal efficiency is verified by measuring the well-known $B\to J/\psi K^\ast$ branching fractions, which are found to be compatible with world averages \cite{pdg}.
As a cross-check, the LFU ratio of ${\cal B}[B\to J/\psi(\to \mu^+ \mu^-) K^\ast]$ and ${\cal B}[B\to J/\psi(\to e^+ e^-) K^\ast]$ 
is measured to be  $1.015\pm 0.025\pm 0.038$, where the first error is statistical and the second  due to uncertainty in   data-MC  corrections for lepton identification.
This cross-check neglects contributions from the $B\to K^\ast\ell\ell$ channel in the $J/\psi$ control region.

The reconstruction procedure for this analysis is optimized for maximal statistical sensitivity to  $R_{K^\ast}$, at the cost of  systematic uncertainties due to the use of multivariate selections in  particle identification.
Systematic uncertainties arises from the determination of the signal yield and reconstruction efficiency.
All considered systematic uncertainties for $R_{K^\ast}$ are listed in \Cref{tab:systtotal}.
The uncertainty due to the signal yield is evaluated by varying the Crystal Ball shape parameters within their uncertainties.
The maximum  yield deviation is taken as the systematic uncertainty. 
The normalizations of  peaking and charmonium  backgrounds   are  varied in the fit  by $\pm 50\%$ and $\pm 25\%$;
these ranges are chosen according to  the maximum uncertainties on the branching fractions used to generate the respective MC samples. 
The resulting signal-yield deviations are included  as part of the systematic uncertainty.
We correct for differences in the lepton-identification efficiency between data and MC by using the results obtained from a control sample of two-photon  $e^+e^-\to e^+e^-\,e^+e^-/e^+e^-\mu^+\mu^-$ events.
The input distributions used by the top-level classifiers are compared between data and simulation, and no  significant differences are found.
In order to estimate the resulting uncertainty, the ratio of $B\to J/\psi K^\ast$ branching fractions between data and MC is obtained in bins of the classifier output.
The obtained ratio is propagated as classifier output-dependent weights to candidates in all fits to \mbc distributions, and changes in the resulting signal yields are taken as systematic uncertainties.
The statistical uncertainty of this reweighting procedure is evaluated in simulations on signal MC samples, and this adds 1-2\% additional uncertainty.
Further uncertainties arise from limited MC statistics.
Effects due to migration of events between different $q^2$ bins are studied using MC events and found to be negligible.
In the  case of  results for the full region of $ q^2>0.045~{\rm GeV}^2/c^4$, the different veto regions for the electron and muon channels need to be accounted for in the determination of  reconstruction efficiency.
This  introduces  model dependence to our signal simulation, which uses form factors from Ref. \cite{Ali:1999mm}. 
We estimate the systematic uncertainty due to this model dependence using different signal MC samples generated with form factors from QCD sum rules \cite{Colangelo:1995jv} and quark models \cite{Melikhov:1997zu}.
The maximum difference in  selection efficiency with respect to the nominal model, in each $q^2$ region, is taken as our estimate for the size of this effect.
This results on average in a difference of $0.4 \pm 2.4\%$ with a maximum of 6.5\%, depending on the mode and $q^2$ region.
As discussed in the beginning, this uncertainty only applies to the branching fractions not to the LFU ratios.
The systematic uncertainty for hadron identification and $K^\ast$ selection is covered in the uncertainty for the  top-level classifiers due to the multivariate selection approach.
For the branching fraction measurements additional uncertainties from tracking (0.35\% per track) and the total number of $B\bar B$ events in data are taken into account. 
The  dominant uncertainty originates from lepton identification, ranging between 5\% and 10\% depending on the mode and $q^2$ region, as also here a more conservative estimation of uncertainty is performed to account for residual correlations with the top-level classifiers.

\begin{table*}
	\scriptsize
	\renewcommand{\arraystretch}{1.2}
	\centering
	\caption{Systematic uncertainties in  $R_{K^\ast}$ for different $q^2$ regions.}
	\label{tab:systtotal}
	\begin{ruledtabular}
		\begin{tabular}{lccccccr}
			$q^2$,  \gevsq      & Signal shape  & Peaking backgrounds  & Charmonium backgrounds  & $e,\mu$ efficiency  & Classifier  & MC size &Total \\
			\hline
			All modes        \\ \hline
			$[0.045, 1.1]$        & 0.025   & 0.026   & 0.001   & 0.027   & 0.030   & 0.006   & 0.054   \\
			$[1.1, 6]$        & 0.033   & 0.070   & 0.013   & 0.065   & 0.038   & 0.008   & 0.109   \\
			$[0.1, 8]$        & 0.002   & 0.054   & 0.051   & 0.058   & 0.024   & 0.005   & 0.098   \\
			$[15, 19]$        & 0.014   & 0.003   & 0.003   & 0.090   & 0.047   & 0.012   & 0.103   \\
			$[0.045, ]$        & 0.008   & 0.031   & 0.023   & 0.061   & 0.026   & 0.004   & 0.077   \\
			\hline
			$B^0$ modes      \\ \hline
			$[0.045, 1.1]$        & 0.005   & 0.049   & 0.001   & 0.024   & 0.112   & 0.007   & 0.125   \\
			$[1.1, 6]$        & 0.062   & 0.070   & 0.012   & 0.082   & 0.062   & 0.010   & 0.140   \\
			$[0.1, 8]$        & 0.019   & 0.033   & 0.018   & 0.058   & 0.049   & 0.006   & 0.087   \\
			$[15, 19]$        & 0.012   & 0.007   & 0.001   & 0.091   & 0.032   & 0.013   & 0.099   \\
			$[0.045, ]$        & 0.018   & 0.031   & 0.021   & 0.073   & 0.033   & 0.006   & 0.090   \\
			\hline
			$B^+$ modes      \\ \hline
			$[0.045, 1.1]$        & 0.060   & 0.006   & 0.000   & 0.033   & 0.060   & 0.013   & 0.092   \\
			$[1.1, 6]$        & 0.060   & 0.086   & 0.009   & 0.045   & 0.092   & 0.010   & 0.147   \\
			$[0.1, 8]$        & 0.040   & 0.048   & 0.107   & 0.060   & 0.023   & 0.010   & 0.140   \\
			$[15, 19]$        & 0.041   & 0.008   & 0.002   & 0.089   & 0.052   & 0.028   & 0.115   \\
			$[0.045, ]$        & 0.018   & 0.025   & 0.023   & 0.044   & 0.015   & 0.005   & 0.061   \\
		\end{tabular}
	\end{ruledtabular}
\end{table*}

\begin{figure}
	\centering
	\subfigure{
		\includegraphics[width=0.4\textwidth]{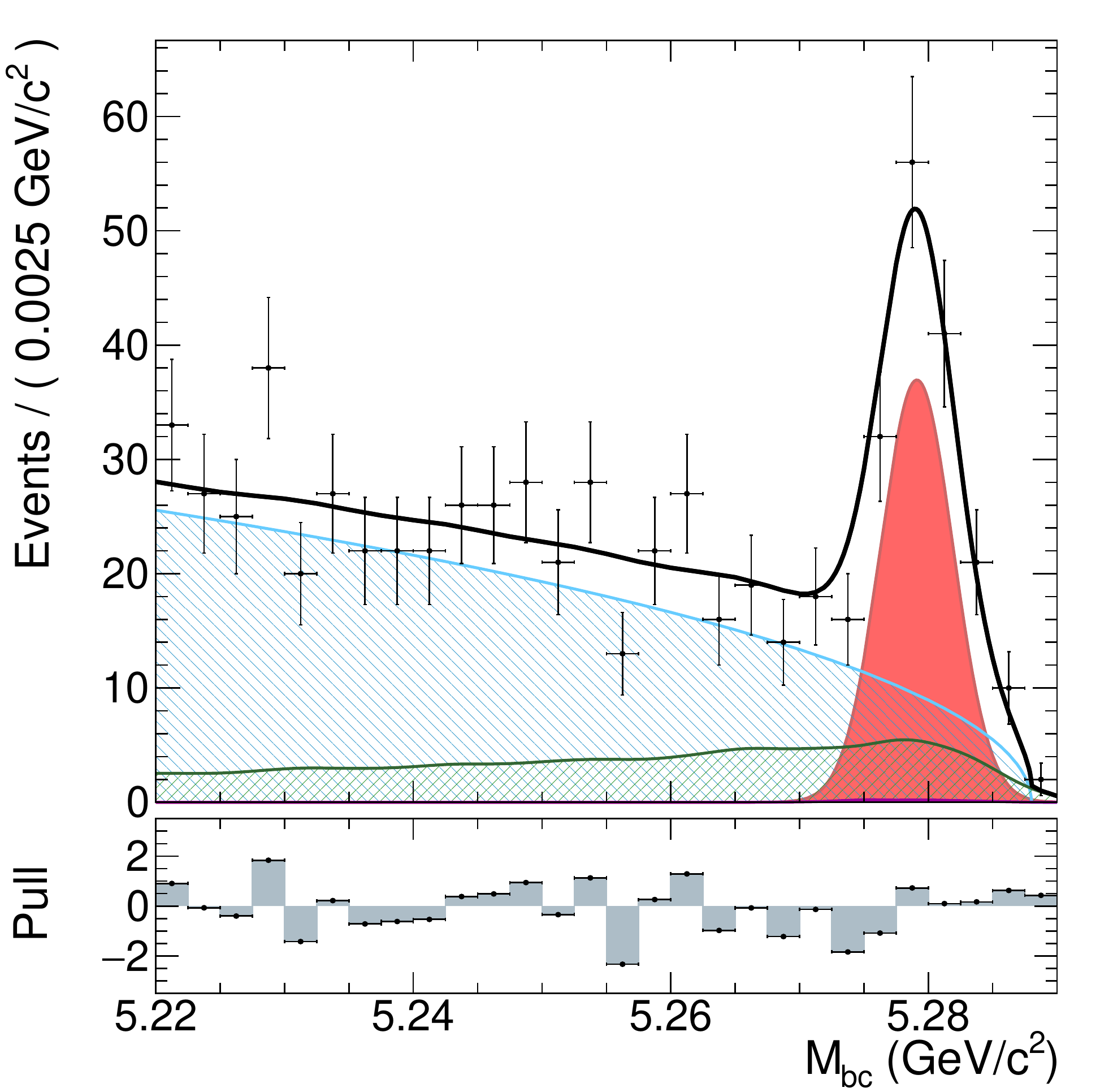}          
	}%
	
	\subfigure{
		\includegraphics[width=0.4\textwidth]{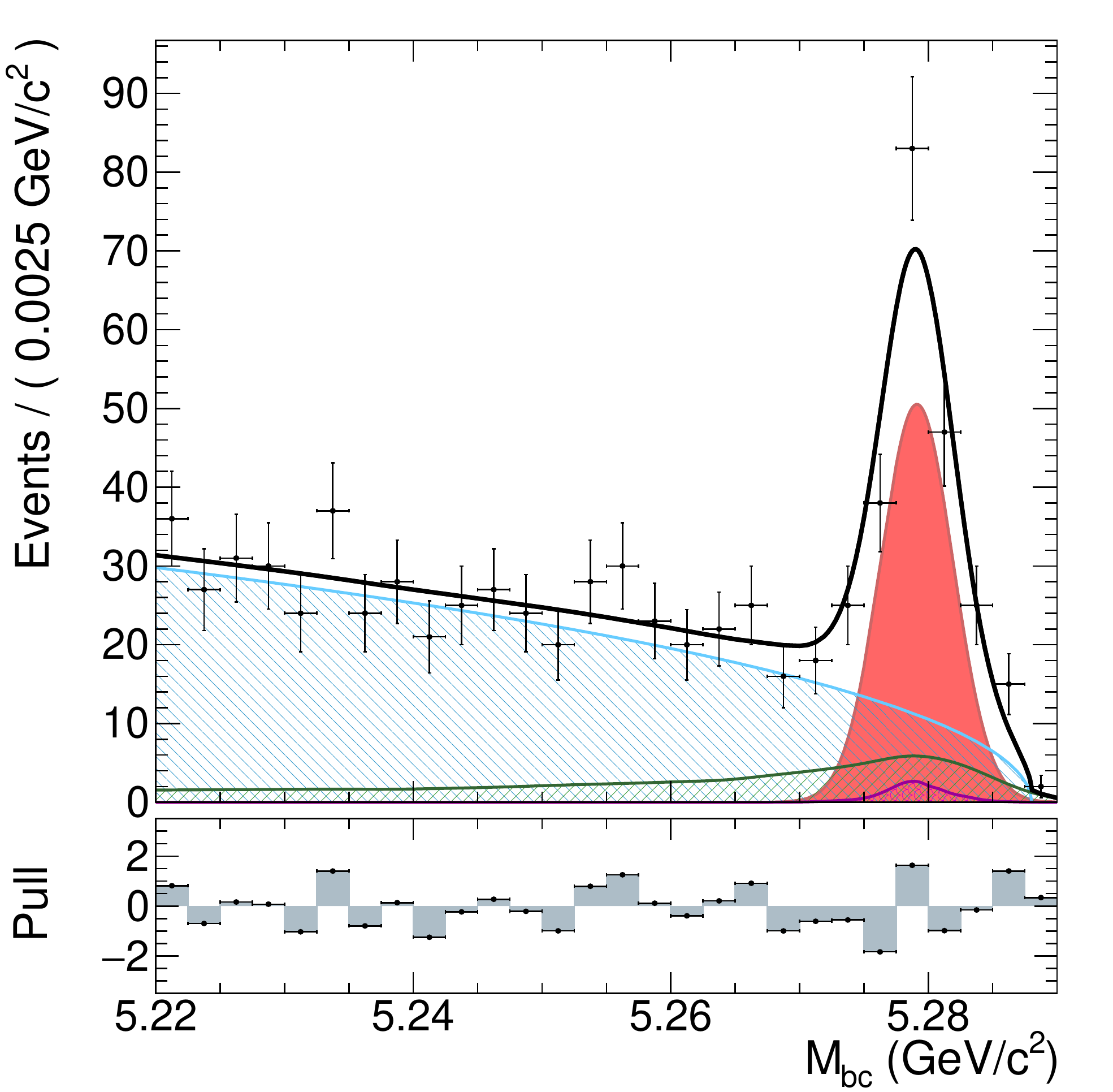}          
	}      
	\caption{Results of the combined $B^+$ and $B^0$ signal yield fit to the $M_{\rm bc}$ distributions for the electron (top) and muon (bottom) modes for $q^2>0.045$ \gevsq. \fitcomponentstwo}
	\label{fig:resultmbc}
\end{figure}

\begin{figure*}
	\centering
	\subfigure[]{
		\includegraphics[width=\factorh\textwidth]{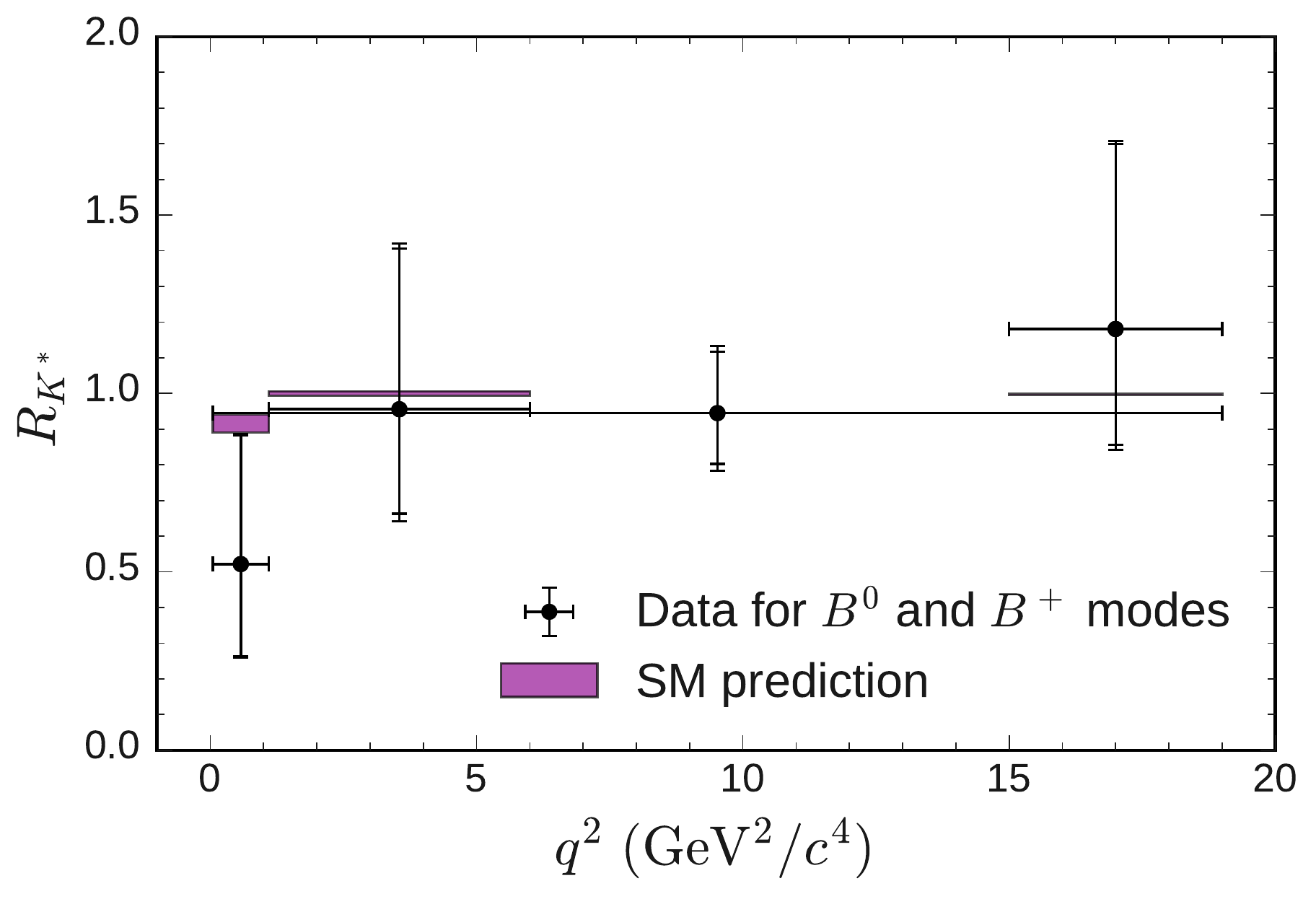}          
	}      	
	\subfigure[]{
		\includegraphics[width=\factorh\textwidth]{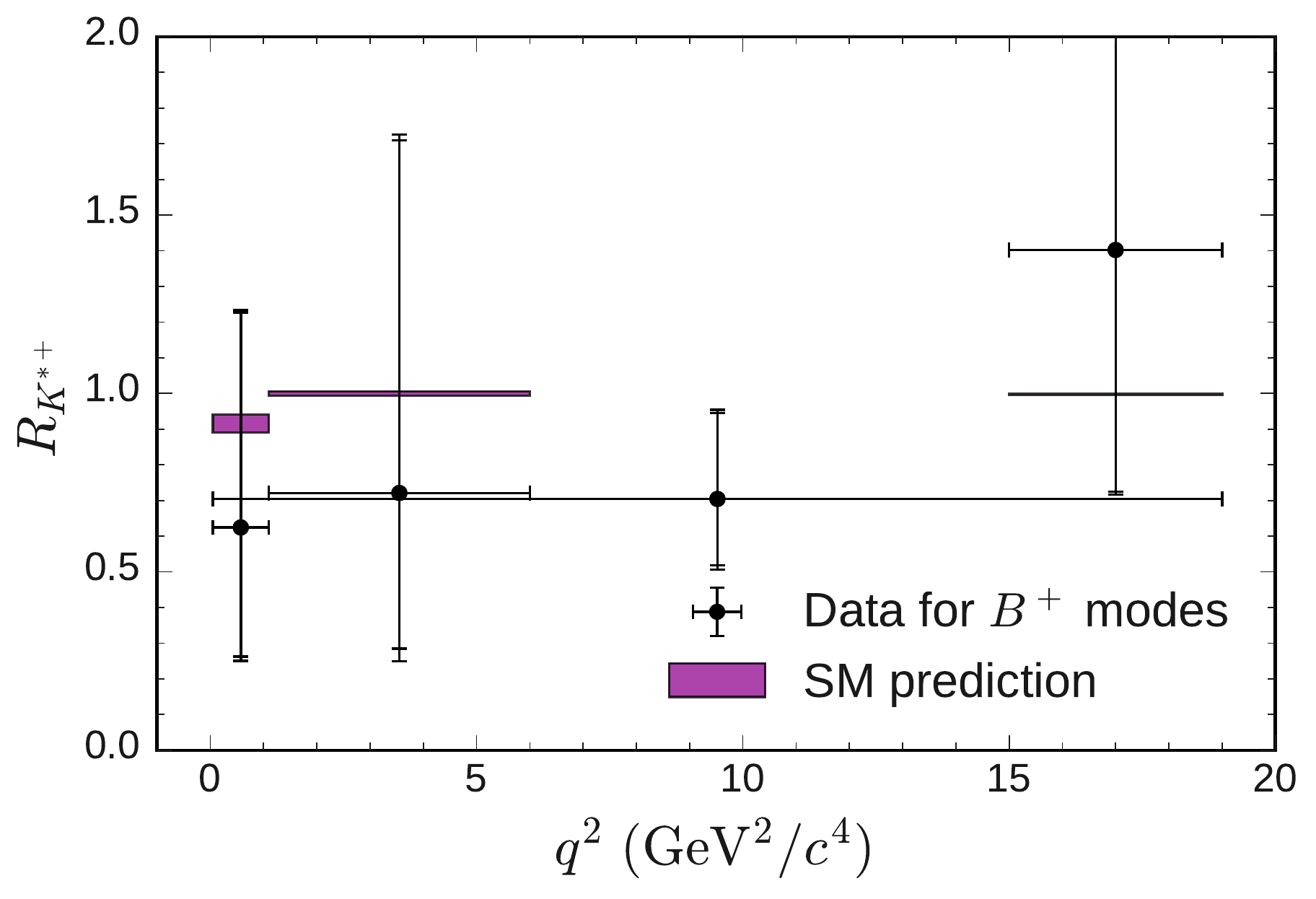}          
	}   
	\subfigure[]{
		\includegraphics[width=\factorh\textwidth]{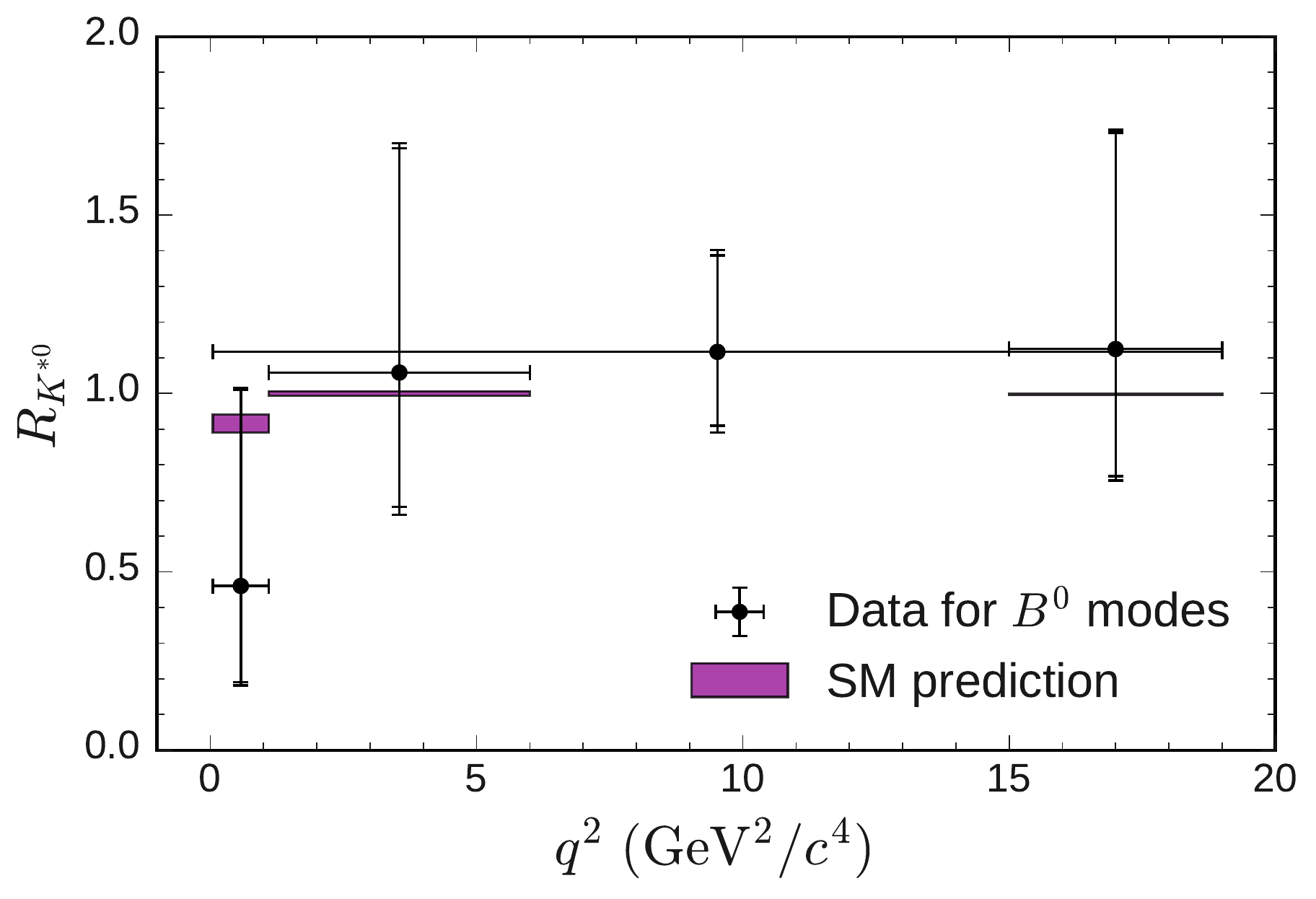}          
	}      
	\caption{Results for $R_{K^\ast}$  compared to SM predictions from  Refs. 
		\cite{DHMVeemumu,Capdevila:2017bsm}.
		The separate vertical error bars indicate the statistical and total uncertainty.
	}
	\label{fig:resultrk}
\end{figure*}

In the range  $q^2>0.045$ \gevsq we find  $103.0^{+13.4}_{-12.7}$ ($139.9^{+16.0}_{-15.4}$) events in the electron (muon) channels.
Example fits are presented in Fig.~\ref{fig:resultmbc}.
Using the fitted signal yields we   construct the LFU ratio $R_{K^\ast}$ for all signal channels combined, as well as separate ratios  for the $B^0$ and $B^+$ decays, $R_{K^{\ast0}}$ and $R_{K^{\ast +}}$.
Our measurement of $R_{K^{\ast+}}$ is the first  ever performed.
Results  are shown in  \Cref{tab:resrkstar} and Fig.~\ref{fig:resultrk}.
\squeezetable
\begin{table}
	\renewcommand{\arraystretch}{1.4}
	\centering
	\caption{Results for $R_{K^\ast}$, $R_{K^{\ast 0}}$, and $R_{K^{\ast +}}$. The first uncertainty  is statistical and the second is    systematic.
	}
	\label{tab:resrkstar}
	
	\begin{ruledtabular}
		\begin{tabular}{lccc}
			$q^2$  in \gevsq & All modes & $B^0$ modes & $B^+$ modes  \\
			\hline
			$[0.045, 1.1]$  	&	 $ 0.52^{+0.36}_{-0.26}\pm0.06 $  	&	 $ 0.46^{+0.55}_{-0.27}\pm0.13 $  	&	 $ 0.62^{+0.60}_{-0.36}\pm0.09 $  \\
			$[1.1, 6]$  	&	 $ 0.96^{+0.45}_{-0.29}\pm0.11 $  	&	 $ 1.06^{+0.63}_{-0.38}\pm0.14 $  	&	 $ 0.72^{+0.99}_{-0.44}\pm0.15 $  \\
			$[0.1, 8]$  	&	 $ 0.90^{+0.27}_{-0.21}\pm0.10 $  	&	 $ 0.86^{+0.33}_{-0.24}\pm0.09 $  	&	 $ 0.96^{+0.56}_{-0.35}\pm0.14 $  \\
			$[15, 19]$  	&	 $ 1.18^{+0.52}_{-0.32}\pm0.11 $  	&	 $ 1.12^{+0.61}_{-0.36}\pm0.10 $  	&	 $ 1.40^{+1.99}_{-0.68}\pm0.12 $  \\
			$[0.045, ]$  	&	 $ 0.94^{+0.17}_{-0.14}\pm0.08 $  	&	 $ 1.12^{+0.27}_{-0.21}\pm0.09 $  	&	 $ 0.70^{+0.24}_{-0.19}\pm0.06 $  \\
		\end{tabular}
	\end{ruledtabular}
\end{table}
The branching fractions are calculated assuming equal production of $B^+$ and $B^0$ mesons  and the results are presented in \Cref{tab:brs}.

\squeezetable
\begin{table*}
	\centering
	\caption{Results for the branching fractions in $[10^{-7}]$ in the corresponding $q^2$ range in \gevsq.
		The first uncertainty is statistical and the second is systematic.
	}
	\label{tab:brs}
	\renewcommand{\arraystretch}{1.2}
	\begin{ruledtabular}
		\begin{tabular}{lrrrr}
			Mode  &$q^2\in [1.1, 6]$   &$q^2\in [0.1, 8]$  &$q^2\in [15, 19]$  &$q^2 > 0.045$ \\
			\hline
			${\cal B}(\Bto{511335})$  & $1.8^{+0.6}_{-0.6}\pm0.2$  & $3.7^{+0.9}_{-0.9}\pm0.4$  & $2.0^{+0.6}_{-0.5}\pm0.2$  & $9.2^{+1.6}_{-1.6}\pm0.8$  \\
			${\cal B}(\Bto{511339})$  & $1.9^{+0.6}_{-0.5}\pm0.3$  & $3.2^{+0.8}_{-0.8}\pm0.4$  & $2.2^{+0.5}_{-0.4}\pm0.2$  & $10.3^{+1.3}_{-1.3}\pm1.1$  \\
			${\cal B}(\Bto{521345})$  & $1.7^{+1.0}_{-1.0}\pm0.2$  & $4.6^{+1.6}_{-1.5}\pm0.7$  & $2.1^{+1.2}_{-1.0}\pm0.2$  & $14.1^{+3.1}_{-2.8}\pm1.8$  \\
			${\cal B}(\Bto{521349})$  & $1.2^{+0.9}_{-0.7}\pm0.2$  & $4.4^{+1.6}_{-1.4}\pm0.5$  & $2.9^{+1.0}_{-0.8}\pm0.3$  & $9.9^{+2.4}_{-2.3}\pm1.1$  \\
		\end{tabular}
	\end{ruledtabular}
\end{table*}

In summary, all our results are consistent  with the SM expectations  \cite{DHMVeemumu,Capdevila:2017bsm}.
Global analyses of measurements of \bsll mediated decays prefer NP models that predict $R_{K^{\ast}}$ values smaller than unity \cite{Capdevila:2017bsm}.
The largest deviation along this direction is observed  in the lowest $q^2$ bin, in the same region where LHCb reports a measurement deviating from the SM \cite{Aaij:2017vbb}.
Our separate results for the $B$-meson isospin partners, $R_{K^{\ast +}}$ and $R_{K^{\ast 0}}$, are statistically compatible, which would also be expected if contributions from  NP arise from the \bsll transition.
The  Belle II experiment \cite{Abe:2010gxa,Kou:2018nap} is expected to record a 50 times larger data sample than Belle, providing ideal conditions to precisely study lepton flavour universality  in these modes.


We thank the KEKB group for the excellent operation of the
accelerator; the KEK cryogenics group for the efficient
operation of the solenoid; and the KEK computer group, and the Pacific Northwest National
Laboratory (PNNL) Environmental Molecular Sciences Laboratory (EMSL)
computing group for strong computing support; and the National
Institute of Informatics, and Science Information NETwork 5 (SINET5) for
valuable network support.  We acknowledge support from
the Ministry of Education, Culture, Sports, Science, and
Technology (MEXT) of Japan, the Japan Society for the 
Promotion of Science (JSPS), and the Tau-Lepton Physics 
Research Center of Nagoya University; 
the Australian Research Council including grants
DP180102629, 
DP170102389, 
DP170102204, 
DP150103061, 
FT130100303; 
Austrian Science Fund (FWF);
the National Natural Science Foundation of China under Contracts
No.~11435013,  
No.~11475187,  
No.~11521505,  
No.~11575017,  
No.~11675166,  
No.~11705209;  
Key Research Program of Frontier Sciences, Chinese Academy of Sciences (CAS), Grant No.~QYZDJ-SSW-SLH011; 
the  CAS Center for Excellence in Particle Physics (CCEPP); 
the Shanghai Pujiang Program under Grant No.~18PJ1401000;  
the Ministry of Education, Youth and Sports of the Czech
Republic under Contract No.~LTT17020;
the Carl Zeiss Foundation, the Deutsche Forschungsgemeinschaft, the
Excellence Cluster Universe, and the VolkswagenStiftung;
the Department of Science and Technology of India; 
the Istituto Nazionale di Fisica Nucleare of Italy; 
National Research Foundation (NRF) of Korea Grant
Nos.~2016R1\-D1A1B\-01010135, 2016R1\-D1A1B\-02012900, 2018R1\-A2B\-3003643,
2018R1\-A6A1A\-06024970, 2018R1\-D1A1B\-07047294, 2019K1\-A3A7A\-09033840,
2019R1\-I1A3A\-01058933;
Radiation Science Research Institute, Foreign Large-size Research Facility Application Supporting project, the Global Science Experimental Data Hub Center of the Korea Institute of Science and Technology Information and KREONET/GLORIAD;
the Polish Ministry of Science and Higher Education and 
the National Science Center;
the Ministry of Science and Higher Education of the Russian Federation, Agreement 14.W03.31.0026; 
University of Tabuk research grants
S-1440-0321, S-0256-1438, and S-0280-1439 (Saudi Arabia);
the Slovenian Research Agency;
Ikerbasque, Basque Foundation for Science, Spain;
the Swiss National Science Foundation; 
the Ministry of Education and the Ministry of Science and Technology of Taiwan;
and the United States Department of Energy and the National Science Foundation.

\end{document}